\documentclass[prc,twocolumn,aps,superscriptaddress,showpacs]{revtex4}
\usepackage{amssymb}
\usepackage{amsmath,bm}
\usepackage{graphicx}
\usepackage[normalem]{ulem}
\usepackage{booktabs}
\usepackage{threeparttable}
\usepackage[dvips]{color}
\usepackage{leftidx}

\renewcommand\sout{\bgroup \color{red} \ULdepth=-.5ex \ULset}

\begin{document}

\title{Constraints on the skewness coefficient of symmetric nuclear matter within the nonlinear relativistic mean field model}

\author{Bao-Jun Cai}
\affiliation{School of Physics and Astronomy and Shanghai Key Laboratory for
Particle Physics and Cosmology, Shanghai Jiao Tong University, Shanghai 200240, China}
\author{Lie-Wen Chen\footnote{%
Corresponding author: lwchen$@$sjtu.edu.cn}} 
\affiliation{School of
Physics and Astronomy and Shanghai Key Laboratory for Particle
Physics and Cosmology, Shanghai Jiao Tong University, Shanghai
200240, China}
\date{\today}

\begin{abstract}
Within the nonlinear relativistic mean field (NL-RMF) model, we show that both the pressure
of symmetric nuclear matter at supra-saturation densities and the maximum mass of
neutron stars are sensitive to the skewness coefficient $J_0$ of symmetric nuclear
matter. Using experimental constraints on the pressure of symmetric nuclear matter
at supra-saturation densities from flow data in heavy ion collisions and the astrophysical
observation of a large mass neutron star PSR J0348+0432, with the former favoring
a smaller $J_0$ while the latter a larger $J_0$, we extract a constraint of
$-494 \mathrm{MeV}\leq J_0\leq -10 \mathrm{MeV}$ based on the NL-RMF model. This constraint is compared
with the results obtained in other analyses.
\end{abstract}


\maketitle

\section{Introduction}

Determination of the equation of state (EOS) of asymmetric nuclear
matter (ANM) is one of fundamental questions in contemporary nuclear
physics and astrophysics. The exact knowledge on the EOS of ANM
provides important information on the in-medium nuclear effective
interactions which play a central role in understanding the
structure and decay properties of finite nuclei as well as the
related dynamical problems in nuclear
reactions\,(\cite{Bla80,LiBA98,Dan02,Bar05,Ste05,Che07a,LCK08,Nat10,Tsa12,Tra12,Che14,Hor14,LiBA14,Liu15,Bal16,LiBA17}).
The EOS of ANM also plays a decisive role in understanding a number
of important issues in astrophysics including the structure and
evolution of neutron stars as well as the mechanism of supernova
explosion\,(\cite{Gle00,Lat04,Lat12,SNRev,Hem12,Mei13,Oer17}).
Conventionally, the EOS of ANM is given by the binding energy per
nucleon as functions of nucleon density $\rho$ and isospin asymmetry
$\delta$, i.e., $E(\rho, \delta)$, and some bulk characteristic
parameters defined at the saturation density $\rho_0$ of symmetric
nuclear matter (SNM) are usually introduced to quantitatively
characterize the EOS of ANM. For example, the energy $E_0(\rho_0)$
and incompressibility $K_0$ of SNM as well as the symmetry energy
$E_{\mathrm{sym}}(\rho_0)$ and its slope parameter $L $ are the four
famous lower-order bulk characteristic parameters of EOS of ANM.
These bulk parameters defined at $\rho_0 $ provide important
information on both sub- and supra-saturation density behaviors of
the EOS of ANM\,(\cite{Che09,Che11a}).

Based on the empirical liquid-drop-like model analyses of high
precision data about nuclear masses, the $E_0(\rho_0)$ is well known
to be about $-16$ MeV. The incompressibility has been determined to
be $K_0 = 240 \pm 40$ MeV from analyzing experimental data of
nuclear giant monopole resonances
(GMR)\,(\cite{Bla80,You99,Shl06,Col09,Pie10,Che12}). For
$E_\text{sym}(\rho_0)$ and $L $, the existing constraints extracted
from terrestrial laboratory measurements and astrophysical
observations are found to be essentially consistent with
$E_{\text{sym}}({\rho _{0}}) = 32.5 \pm 2.5$ MeV and $L = 55 \pm 25$
MeV (see, e.g., Refs.~\cite{Che12a,Li12}). While these lower-order bulk
characteristic parameters have been relatively well determined or in
significant progress, our knowledge on the higher-order bulk
characteristic parameters remains very limited. Following
$E_0(\rho_0)$, $K_0$, $E_\text{sym}(\rho_0)$ and $L $, the next bulk
characteristic parameter should be the skewness coefficient $J_0$
(also denoted as $K'$ or $Q_0$ in some literature) of SNM, which is
related to the third-order density derivative of the binding energy
per nucleon of SNM at $\rho_0 $. The higher-order bulk
characteristic parameter $J_0$ is expected to be important for the
high density behaviors of nuclear matter EOS and thus may play an
essential role in heavy ion collisions (HIC), the structure and
evolution of neutron stars, supernova explosion, and gravitational
wave radiation from merging of compact stars. To our best knowledge,
so far there is very little experimental information on the $J_0$
parameter, and it is thus of great interest and critical importance
to constrain the $J_0$ parameter, which is the main motivation of
the present work.

Within the nonlinear relativistic
mean field (RMF) model, we demonstrate in this work that the
pressure of SNM at supra-saturation densities and the maximum mass
of neutron stars provide good probes of the skewness coefficient
$J_0$. In particular, combining the experimental constraints
on the pressure of SNM at supra-saturation densities from flow data
in HIC and the recent astrophysical observation of a large mass neutron
star PSR J0348+0432, one can obtain a strong constraint on the $J_0$
parameter.

\section{The skewness coefficient $J_0$ in nonlinear RMF model}

\subsection{Nuclear matter characteristic parameters}

The EOS of isospin asymmetric nuclear matter, namely $E(\rho,
\delta)$, can be expanded as a power series of even-order terms in
$\delta $ as
\begin{equation}
E(\rho ,\delta )\simeq E_{0}(\rho )+E_{\text{sym}}(\rho )\delta ^{2}+%
\mathcal{O}(\delta ^{4}), \label{EoSpert}
\end{equation}%
where $E_{0}(\rho )=E(\rho ,\delta =0)$ is the EOS of symmetric
nuclear matter, and the symmetry energy is expressed as
\begin{equation}
E_{\text{sym}}(\rho )= \left. \frac{1}{2}\frac{\partial ^{2}E(\rho
,\delta )}{\partial \delta ^{2}}\right\vert _{\delta =0}.
\label{DefEsym}
\end{equation}%
Around the saturation density $\rho _{0}$, the $E_{0}(\rho )$ can be
expanded, e.g., up to $3$rd-order in density, as,
\begin{equation}
E_{0}(\rho )=E_{0}(\rho _{0})+\frac{K_{0}}{2!}\chi
^{2}+\frac{J_{0}}{3!}\chi ^{3}+\mathcal{O}(\chi ^{4}),
\label{DenExp0}
\end{equation}%
where $\chi =(\rho -\rho _{0})/3\rho _{0} $ is a dimensionless
variable characterizing the deviations of the density from the
saturation density $\rho _{0}$. The first term $E_{0}(\rho _{0})$ on
the right-hand-side of Eq. (\ref{DenExp0}) is the binding energy per
nucleon in SNM at $\rho _{0}$ and the coefficients of other terms
are
\begin{align}
K_{0} =&\left. 9\rho _{0}^{2}\frac{\text{d} ^{2}E_{0}(\rho
)}{\text{d} \rho
^{2}}\right\vert _{\rho =\rho _{0}},~~  \label{K0} \\
J_{0} =&\left. 27\rho _{0}^{3}\frac{\text{d} ^{3}E_{0}(\rho
)}{\text{d} \rho ^{3}}\right\vert _{\rho =\rho _{0}},  \label{J0}
\end{align}%
where $K_{0}$ is the well-known incompressibility coefficient of SNM
and $J_{0}$ is the skewness coefficient of SNM, i.e. the $3$rd-order
incompressibility coefficient of SNM\,(\cite{Che09,Che11a}).

Similarly, one can expand the $E_{\mathrm{sym}}(\rho )$ around an
arbitrary reference density $\rho _{\text{r}}$ as
\begin{equation}
E_{\text{sym}}(\rho )=E_{\text{sym}}({\rho _{\text{r}}})+L(\rho
_{\text{r}})\chi_{\text{r}}+\mathcal{O}(\chi_{\text{r}} ^{2}),
\label{EsymLKr}
\end{equation}
with $\chi_{\text{r}}=(\rho -\rho _{\text{{r}}})/3\rho _{\text{r}}$,
and the slope parameter of the symmetry energy at $\rho _{\text{r}}$
is expressed as~\cite{Zha13}
\begin{align}
L(\rho_{\text{r}})=&\left.3\rho_{\text{r}}
\frac{\text{d}E_{\mathrm{sym}}(\rho )}{\text{d}\rho } \right|_{\rho
=\rho_{\text{r}} }.
\end{align}
For $\rho_{\text{r}} = \rho_0$, the $L(\rho_{\text{r}})$ is reduced
to the conventional slope parameter $L\equiv 3\rho_0
{\text{d}E_{\mathrm{sym}}(\rho )}/{\text{d}\rho }|_{\rho =\rho_0}$.

If $\delta$ and $\chi$ are assumed to be small quantities on the
same order, nuclear matter bulk characteristic parameters can then
be classified accordingly in different orders. For example, $L$ and
$J_0$ are on the same order-$3$, i.e., $\delta^2\chi$ for $L$ and
$\delta^0\chi^3$ for $J_0$. In this sense, $E_{0}(\rho _{0})$ is on
the order-$0$, $K_0$ and $E_{\text{sym}}({\rho _{0}})$ are on the
order-$2$. To see the role of $J_0$ in the EOS of SNM, one can
re-write Eq.~(\ref{DenExp0}) in a slightly different form as
\begin{equation}
E_0(\rho)\simeq E_0(\rho_0)+\frac{1}{2}K_0\chi^2\left(1+\frac{\chi
J_0}{3K_0}\right).
\end{equation}
Assuming $J_0$ has roughly the same magnitude as $K_0$, one can see
that the contribution from the $J_0$ term to the EOS of SNM becomes
comparable with that from the $K_0$ term if the baryon density is
larger than about $3\rho_0$, corresponding to the typical densities
inside a neutron star. On the other hand, the $J_0$ term plays a
minor role for the EOS of SNM at subsaturation densities relevant
for nuclear structure properties. As we will see later, the pressure
of SNM at supra-saturation densities and the maximum mass of neutron
stars indeed display strong sensitivity on the $J_0$ parameter.

\subsection{Nuclear matter characteristic parameters in nonlinear RMF model}

The nonlinear RMF model has made great success during the last
decades in describing many nuclear phenomena (see, e.g.,
\cite{Ser86,Rei89,Rin96,Men06,Sug94,Ren96,Lal97,Lon04,Jia05,Jia10,Fat10,Agr12,Fat13}).
In the following, we briefly describe the nonlinear RMF model that
we shall adopt in this work and present some useful expressions of
nuclear matter characteristic parameters, especially the skewness
coefficient $J_0$. The interacting Lagrangian of the nonlinear RMF
model supplemented with couplings between the isoscalar and the
isovector mesons
reads\,(\cite{Mul96,Hor01,Tod05,Che07,Cai12})
\begin{align}
\mathcal{L}=&\overline{\psi}\left[\gamma_{\mu}(i\partial^{\mu}-g_{\omega}\omega^{\mu}
-g_{\rho}\vec{\mkern1mu\rho}^{\mu}\cdot\vec{\mkern1mu\tau})-(M-g_{\sigma}\sigma)\right]\psi\notag\\
&-\frac{1}{2}m_{\sigma}^2\sigma^2+\frac{1}{2}\partial_{\mu}\sigma\partial^{\mu}\sigma
-U(\sigma)\notag\\
&+\frac{1}{2}m_{\omega}^2\omega_{\mu}\omega^{\mu}-\frac{1}{4}\omega_{\mu\nu}\omega^{\mu\nu}
+\frac{1}{4}c_{\omega}\left(g_{\omega}\omega_{\mu}\omega^{\mu}\right)^2\notag\\
&+\frac{1}{2}m_{\rho}^2\vec{\mkern1mu\rho}_{\mu}\cdot\vec{\mkern1mu\rho}^{\mu}
-\frac{1}{4}\vec{\mkern1mu\rho}_{\mu\nu}\cdot\vec{\mkern1mu\rho}^{\mu\nu}\notag\\
&+\frac{1}{2}g_{\rho}^2\vec{\mkern1mu\rho}_{\mu}\cdot\vec{\mkern1mu\rho}^{\mu}
\Lambda_{\mathrm{V}}g_{\omega}^2\omega_{\mu}\omega^{\mu},\label{NLRMF}
\end{align}%
where $\omega_{\mu \nu }\equiv \partial _{\mu }\omega _{\nu
}-\partial _{\nu
}\omega _{\mu }$ and$~\rho_{\mu \nu }\equiv \partial _{\mu }\vec{%
\mkern1mu\rho }_{\nu }-\partial _{\nu }\vec{\mkern1mu\rho }_{\mu }$
are strength tensors for $\omega $ field and $\rho $ field,
respectively. $\psi $, $\sigma $, $\omega _{\mu }$,
$\vec{\mkern1mu\rho }_{\mu }$ are nucleon field, isoscalar-scalar
field, isoscalar-vector field and isovector-vector field,
respectively, and the arrows denote the vector in isospin space,
$U({\sigma})=b_{\sigma}M(g_{\sigma}\sigma)^3/3+c_{\sigma}(g_{\sigma}\sigma)^4/4
$ is the self interaction term for $\sigma$ field. $\Lambda
_{\textrm{V}}$ represents the coupling constant between the
isovector $\rho $ meson and the isoscalar $\omega $ meson and it is
important for the description of the density dependence of the
symmetry energy. In addition, $M$ is the nucleon mass and $m_{\sigma
}$, $m_{\omega }$, $m_{\rho }$ are masses of mesons.

In the mean field approximation, after neglecting effects of
fluctuation and correlation, meson fields are replaced by their
expectation values, i.e., $\overline{\sigma}\rightarrow \sigma $,
$\overline{\omega}_{0}\rightarrow \omega _{\mu } $,
$\overline{\rho}_{0}^{(3)}\rightarrow \vec{\mkern1mu\rho }_{\mu }$,
where subscript \textquotedblleft $0$" indicates zeroth component of
the four-vector, superscript \textquotedblleft ($3$)" indicates
third component of the isospin. Furthermore, we also use in this
work the non-sea approximation which neglects the effect due to
negative energy states in the
Dirac sea. The mean field equations are then expressed as 
\begin{align}
m_{\sigma }^{2}\overline{\sigma}=& g_{\sigma }\left[ \rho
_{\textrm{S}}-b_{\sigma }M\left(
g_{\sigma }\overline{\sigma}\right) ^{2}-c_{\sigma }\left( g_{\sigma }\overline{\sigma}%
\right) ^{3}\right], \\
m_{\omega }^{2}\overline{\omega}_{0}=& g_{\omega }\left[ \rho
-c_{\omega }\left(
g_{\omega }\overline{\omega}_{0}\right) ^{3}-\Lambda _{\textrm{V}}g_{\omega }\overline{\omega}%
_{0}\left( g_{\rho }\overline{\rho}_{0}^{(3)}\right) ^{2}\right], \\
m_{\rho }^{2}\overline{\rho}_{0}^{(3)}=& g_{\rho }\left[ \rho
_{\textrm{p}}-\rho
_{\textrm{n}}-\Lambda _{\textrm{V}}g_{\rho }\overline{\rho}_{0}^{(3)}\left( g_{\omega }\overline{%
\omega}_{0}\right) ^{2}\right],
\end{align}%
where
\begin{equation}
\rho =\langle \overline{\psi}\gamma ^{0}\psi \rangle =\rho
_{\textrm{n}}+\rho _{\textrm{p}},~~\rho _{\textrm{S}}=\langle
\overline{\psi}\psi \rangle =\rho _{\textrm{S,n}}+\rho
_{\textrm{S,p}}, \label{density}
\end{equation}%
are the baryon density and scalar density, respectively, with the
latter given by
\begin{align}
\rho _{\textrm{S},J} &=\frac{2}{(2\pi )^{3}}\int_{0}^{k_{\textrm{F}}^{J}}\textrm{d}\textbf{k}\frac{%
M^{\ast }}{\sqrt{|\textbf{k}|^{2}+{M^{\ast }}^{2}}}  \notag \\
&=\frac{M^{\ast }}{2\pi ^{2}}\left[ k_{\textrm{F}}^{J}\mathcal{E}_{\textrm{F}}^{J\ast }-{%
M^{\ast
2}}\ln\left(\frac{k_{\textrm{F}}^{J}+\mathcal{E}_{\textrm{F}}^{J\ast
}}{M^{\ast }}\right)\right] ,J=\textrm{p,n}.  \label{ScaDen}
\end{align}%
In the above expression, we have $\mathcal{E}_{\textrm{F}}^{J\ast }=\sqrt{k_{\textrm{F}}^{J 2}+M^{%
\ast 2}}$ and the nucleon Dirac mass is defined as
\begin{align}
M^{\ast }\equiv M_{\textrm{dirac}}^{\ast}=M-g_{\sigma
}\overline{\sigma}.
\end{align}
$k_{\textrm{F}}^{J}=k_{\textrm{F}}(1+\tau _{3}^J\delta )^{1/3}$ is
the Fermi momentum with $\tau _{3}^{\textrm{n}}=+1$ for neutrons and
$\tau _{3}^{\textrm{p}}=-1$ for protons, and $k_{\textrm{F}}=(3\pi
^{2}\rho /2)^{1/3}$ is the Fermi momentum for SNM at $\rho $.

The energy-momentum density tensor for the interacting Lagrangian
density in Eq. (\ref{NLRMF}) can be written as
\begin{equation}
\mathcal{T}^{\mu \nu }=\overline{\psi}i\gamma ^{\mu }\partial ^{\nu
}\psi +\partial ^{\mu }\sigma \partial ^{\nu }\sigma -\omega^{\mu
\eta }\partial ^{\nu
}\omega _{\eta }-\vec{\mkern1mu \rho }^{\mu \eta }\partial ^{\nu }\vec{\mkern%
1mu\rho }_{\eta }-\mathcal{L}g^{\mu \nu },  \label{EnMomTen}
\end{equation}%
where $g_{\mu \nu }=(+,-,-,-)$ is the Minkowski metric. In the mean
field approximation, the mean value of time (zero) component of the
energy-momentum density tensor is the energy density of the nuclear
matter
system, i.e.,%
\begin{align}
\varepsilon =& \langle \mathcal{T}^{00}\rangle \notag\\
=&\varepsilon _{\mathrm{kin}%
}^{\textrm{n}}+\varepsilon _{\mathrm{kin}}^{\textrm{p}}+\frac{1}{2}\left[ m_{\sigma }^{2}\overline{%
\sigma}^{2}+m_{\omega }^{2}\overline{\omega}_{0}^{2}+m_{\rho }^{2}\left( \overline{\rho%
}_{0}^{(3)}\right) ^{2}\right]  \notag \\
& +\frac{1}{3}b_{\sigma }(g_{\sigma
}\overline{\sigma})^{3}+\frac{1}{4}c_{\sigma
}(g_{\sigma }\overline{\sigma})^{4}+\frac{3}{4}c_{\omega }(g_{\omega }\overline{\omega}%
_{0})^{4}  \notag \\
& +\frac{3}{2}\left( g_{\rho }\overline{\rho}_{0}^{(3)}\right) ^{2}\Lambda _{\textrm{V}}(g_{\omega }\overline{\omega}%
_{0})^{2},
\end{align}%
where%
\begin{align}
\varepsilon _{\mathrm{kin}}^{J} =&\frac{2}{(2\pi )^{3}}\int_{0}^{k_{\textrm{F}}^{J}}\textrm{d}%
\textbf{k}\sqrt{|\textbf{k}|^{2}+{M^{\ast 2}}}  \notag \\
=&\frac{1}{\pi
^{2}}\int_{0}^{k_{\textrm{F}}^{J}}k^{2}\textrm{d}k\sqrt{k^{2}+{M^{\ast
2}}}
\notag \\
=&\frac{1}{4}\left[ 3\mathcal{E}_{\textrm{F}}^{J\ast }\rho
_{J}+M^{\ast }\rho _{\textrm{S},J}\right] ,~~J=\textrm{p,n},
\label{EnDenKin}
\end{align}%
is the kinetic part of the energy density. Similarly, the mean value
of space components of the energy-momentum density tensor
corresponds to the
pressure of the system, i.e.,%
\begin{align}
P=& \frac{1}{3}\sum_{j=1}^{3}\langle \mathcal{T}^{jj}\rangle =P_{\mathrm{kin}%
}^{\textrm{n}}+P_{\mathrm{kin}}^{\textrm{p}}  \notag \\
& -\frac{1}{2}\left[ m_{\sigma }^{2}\overline{\sigma}^{2}-m_{\omega }^{2}\overline{%
\omega}_{0}^{2}-m_{\rho }^{2}\left( \overline{\rho}_{0}^{(3)}\right)
^{2}\right]
\notag \\
& -\frac{1}{3}b_{\sigma }(g_{\sigma
}\overline{\sigma})^{3}-\frac{1}{4}c_{\sigma
}(g_{\sigma }\overline{\sigma})^{4}+\frac{1}{4}c_{\omega }(g_{\omega }\overline{\omega}%
_{0})^{4}  \notag \\
& +\frac{1}{2}\left( g_{\rho }\overline{\rho}_{0}^{(3)}\right) ^{2}\Lambda _{\textrm{V}}(g_{\omega }\overline{\omega}%
_{0})^{2} ,
\end{align}%
where the kinetic part of pressure is given by
\begin{equation}
P_{\mathrm{kin}}^{J}=\frac{1}{3\pi ^{2}}\int_{0}^{k_{\textrm{F}}^{J}}\textrm{d}k\frac{k^{4}}{%
\sqrt{k^{2}+{M^{\ast}}^{2}}},~~J=\textrm{p,n}. \label{pressureKin}
\end{equation}

The EOS of ANM can be calculated through the energy density
$\varepsilon(\rho,\delta)$ by
\begin{equation}
E(\rho ,\delta )=\frac{\varepsilon (\rho ,\delta )}{\rho }-M.
\label{SiEn}
\end{equation}%
The EOS of SNM is just $E_0(\rho)\equiv E(\rho,\delta=0)$, and the
characteristic parameters $K_0$ and $J_0$ can be obtained from the
following expressions
\begin{align}
K_0(\rho)\equiv&9\rho^2\frac{\text{d}^2E_0}{\text{d}\rho^2}=-\frac{9\rho
g_{\sigma}^2M^{\ast 2}}{Q_{\sigma}\mathcal{E}_{\textrm{F}}^{\ast
2}}+\frac{9\rho g_{\omega}^2}{Q_{\omega}}+
\frac{3k_{\textrm{F}}^2}{\mathcal{E}_{\textrm{F}}^{\ast}}-6L_0(\rho),\label{K0_for}\\
J_0(\rho)\equiv&27\rho^3\frac{\text{d}^3E_0}{\text{d}\rho^3}=-\frac{3k_{\textrm{F}}^2}{\mathcal{E}_{\textrm{F}}^{\ast}}-\frac{3k_{\textrm{F}}^4}{\mathcal{E}_{\textrm{F}}^{\ast
3}}
+\frac{27g_{\sigma}^2M^{\ast 2}\rho^2}{Q_{\sigma}\mathcal{E}_{\textrm{F}}^{\ast 3}}\times\notag\\
&\left(\frac{3\pi^2}{2k_{\textrm{F}}\mathcal{E}_{\textrm{F}}^{\ast}}+\frac{2g_{\sigma}^2}{Q_{\sigma}}-\frac{g_{\sigma}M^{\ast}\eta}{Q_{\sigma}^2}
-\frac{2g_{\sigma}^2M^{\ast 2}}{Q_{\sigma}\mathcal{E}_{\textrm{F}}^{\ast 2}}+\frac{\mathcal{E}_{\textrm{F}}^{\ast}\phi}{2Q_{\sigma}}\right)\notag\\
&-\frac{162c_{\omega}g_{\omega}^7\overline{\omega}_0\rho^2}{Q_{\omega}^3}-9K_0(\rho)
,\label{J0_for}
\end{align}
with
\begin{align}L_0(\rho)\equiv&3\rho\frac{\text{d}
E_0}{\text{d}\rho}=3\Bigg[\frac{\mathcal{E}_{\rm{F}}^{\ast}}{4}
-\frac{M^{\ast}\rho_{\rm{S}}}{4\rho}+g_{\omega}\overline{\omega}_0\notag\\
&-\frac{1}{\rho}\left(\frac{1}{2}m_{\sigma}^2\overline{\sigma}^2+U(\overline{\sigma})+\frac{1}{2}m_{\omega}^2\overline{\omega}_0^2+\frac{3}{4}c_{\omega}g_{\omega}^4\overline{\omega}_0^4\right)\Bigg]
.
\end{align}
In the above expressions, we have
\begin{align}
Q_{\sigma }=& m_{\sigma }^{2}+g_{\sigma }^{2}\left( \frac{3\rho _{\textrm{S}}}{{%
M^{\ast }}}-\frac{3\rho }{\mathcal{E}_{\textrm{F}}^{\ast }}\right)
+2b_{\sigma }Mg_{\sigma
}^{3}\overline{\sigma}+3c_{\sigma }g_{\sigma }^{4}\overline{\sigma}^{2},  \label{Qf} \\
Q_{\omega }=& m_{\omega }^{2}+3c_{\omega }g_{\omega }^{4}\overline{\omega}%
_{0}^{2},  \label{Qw}
\end{align}%
and
\begin{align}
\eta=&3g_{\sigma}^3\left(\frac{2\rho_{\textrm{S}}}{M^{\ast
2}}-\frac{3\rho}{M^{\ast}\mathcal{E}_{\textrm{F}}^{\ast}}
+\frac{M^{\ast}\rho}{\mathcal{E}_{\textrm{F}}^{\ast 3}}\right)-2b_{\sigma}Mg_{\sigma}^3-6c_{\sigma}g_{\sigma}^4\overline{\sigma},\label{def_chi}\\
\phi=&\frac{2g_{\sigma}^2k_{\textrm{F}}^2}{\mathcal{E}_{\textrm{F}}^{\ast
3}},\label{def_phi}\end{align} with
$\mathcal{E}_{\textrm{F}}^{\ast}=(k_{\rm{F}}^{2}+M^{\ast 2})^{1/2}$.
To our best knowledge, Eq.~(\ref{J0_for}) gives, for the first time~\cite{Cai14},
the analytical expression of the $J_0$ parameter in the nonlinear
RMF model. In addition, we would like to point out that the general
expression for $E_{\rm{sym}}(\rho)$ and $L(\rho)$ in nonlinear RMF
model has been derived in~\cite{Cai12a}.

\section{Results and discussions}

For the Lagrangian in Eq.~(\ref{NLRMF}), the properties of infinite
nuclear matter is uniquely determined by
$f_{\sigma}=g_{\sigma}/m_{\sigma}$, $b_{\sigma}$, $c_{\sigma}$,
$f_{\omega}=g_{\omega}/m_{\omega}$, $c_{\omega}$,
$f_{\rho}=g_{\rho}/m_{\rho}$, $\Lambda_{\mathrm{V}}$, and $M$. If
the nucleon mass in vacuum is set to be $M = 939$ MeV, one then has
totally seven parameters to determine the properties of infinite
nuclear matter in the nonlinear RMF model. Following the correlation
analysis method proposed in~\cite{Che10} within the
Skyrme-Hartree-Fock (SHF) approach, instead of using directly the
seven microscopic parameters, i.e., $f_{\sigma}$, $b_{\sigma}$,
$c_{\sigma}$, $f_{\omega}$, $c_{\omega}$, $f_{\rho}$,
$\Lambda_{\mathrm{V}}$, one can determine their values explicitly in
terms of seven macroscopic quantities, i.e., $\rho _{0}$,
$E_{0}(\rho_{0})$, $K_{0}$, $J_{0}$, $M_{\mathrm{dirac}}^{\ast
0}\equiv M_{\mathrm{dirac}}^{\ast}(\rho_{0})$,
$E_{\rm{sym}}(\rho_{\mathrm{c}})$, and $L(\rho_{\mathrm{c}})$ where
$\rho_{\mathrm{c}}$ is the cross density whose value is fixed in
this work at $0.11\,\mathrm{fm}^{-3}$\,(\cite{Zha13}). Then, by
varying individually these macroscopic quantities within their known
ranges, one can examine transparently the correlation of
nuclear matter properties with each individual macroscopic quantity.
Recently, this simple correlation analysis method has been successfully
applied to study the neutron skin\,(\cite{Che10,Zha13}) and giant
resonances of finite nuclei\,(\cite{Che12,Zha14}), the higher-order bulk characteristic
parameters of ANM\,(\cite{Che11a}), and the relationship between the
nuclear matter symmetry energy and the symmetry energy coefficient
in the mass formula\,(\cite{Che11}). We would like to point out
although the seven macroscopic quantities defined above coherently act on the
maximum mass of neutron stars and the pressure of the SNM, they are independent with
each other in our analysis since we vary one quantity by keeping other six quantities fixed.
This is one of the main advantages of our approach since the physics of these
macroscopic quantities is different.

\begin{figure*}
\begin{center}
\includegraphics[width=0.7\textwidth]{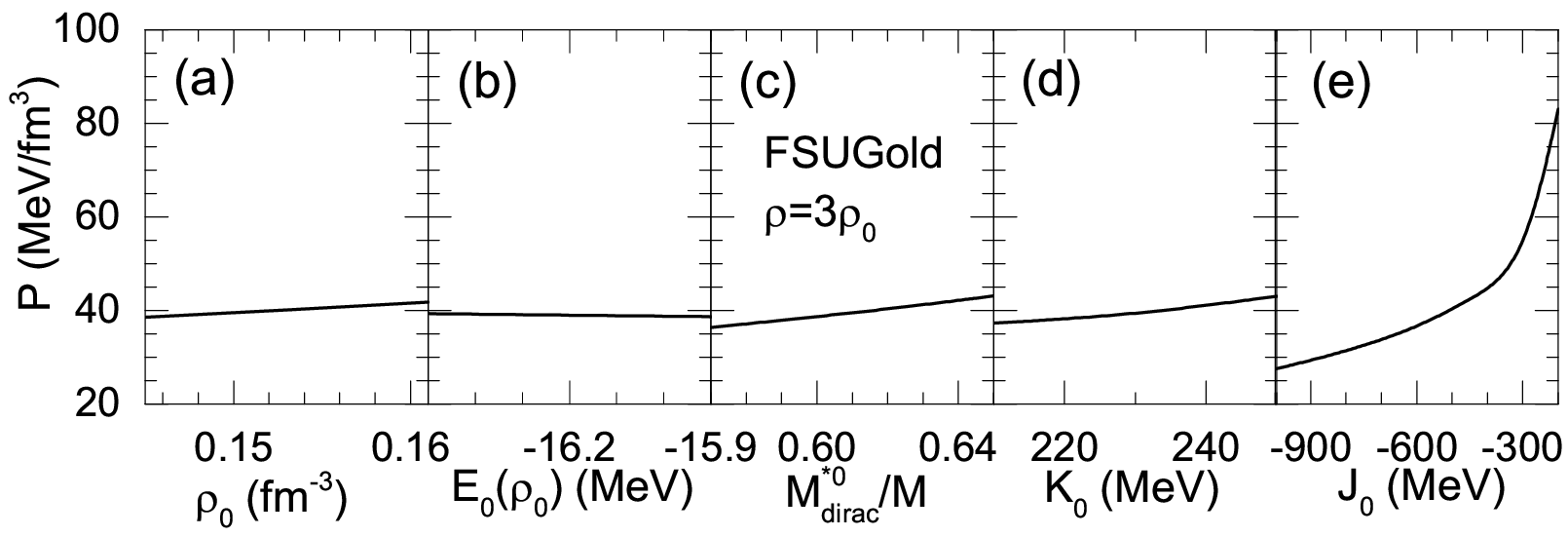}
\caption{Pressure of SNM at $\rho=3\rho_0$ from the nonlinear RMF model based on
the FSUGold interaction by varying individually $\rho_{0}$ (a), $E_{0}(\rho_{0})$ (b),
$M_{\mathrm{dirac}}^{\ast 0}$ (c), $K_{0}$ (d), and $J_{0}$ (e).}
\label{PressCorr}
\end{center}
\end{figure*}

\begin{figure*}
\begin{center}
\includegraphics[width=0.95\textwidth]{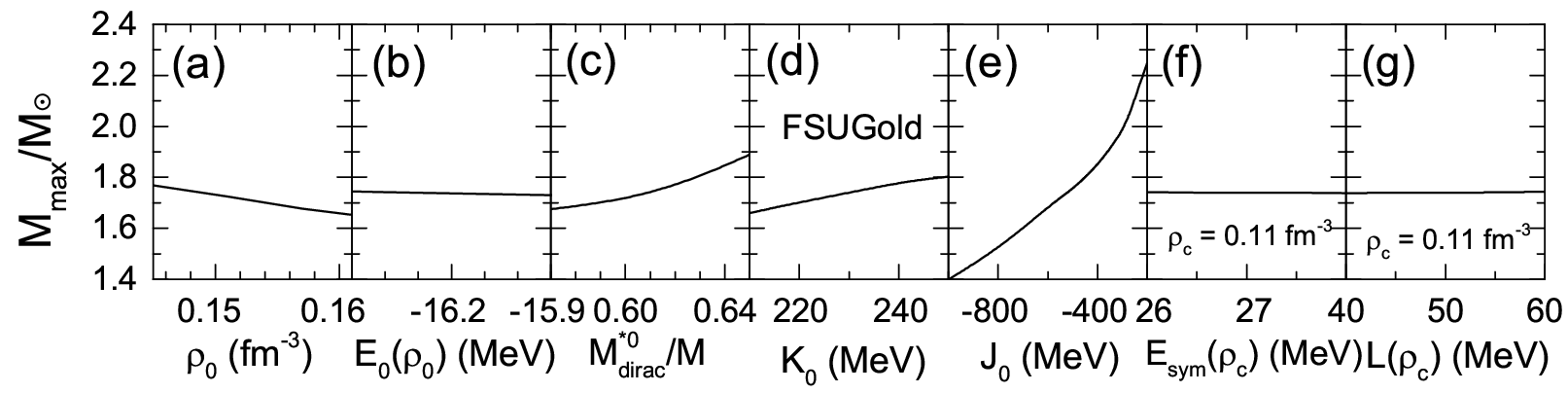}
\caption{\label{MassCorr}  (Color online) Maximum mass of static
neutron stars from the nonlinear RMF model based on the FSUGold
interaction by varying individually $\rho _{0}$ (a),
$E_{0}(\rho_{0})$ (b), $M_{\mathrm{dirac}}^{\ast 0}$ (c),  $K_{0}$
(d), $J_{0}$ (e), $E_{\rm{sym}}(\rho_{\mathrm{c}})$ (f), and
$L(\rho_{\mathrm{c}})$ (g).}
\end{center}
\end{figure*}

To examine the correlation of pressure of SNM at supra-saturation
densities with each macroscopic quantity, we show in
Fig.\,\ref{PressCorr} the pressure of SNM $P(\rho)$ at
$\rho=3\rho_0$ from the nonlinear RMF model based on the FSUGold
interaction\,(\cite{Tod05}) by varying individually $\rho_{0}$,
$E_{0}(\rho_{0})$, $M_{\mathrm{dirac}}^{\ast 0}$, $K_{0}$, and
$J_{0}$ within their empirical uncertain ranges, namely, varying one
quantity at a time while keeping all others at their default values
in FSUGold for which we have $\rho _{0}=0.148$\,fm$^{-3}$,
$E_{0}(\rho_{0})=-16.3$\,MeV, $M_{\mathrm{dirac}}^{\ast 0}=0.61M$,
$K_{0}=230$\,MeV, $J_{0}=-522.6$\,MeV,
$E_{\rm{sym}}(\rho_{\mathrm{c}})=27.11$\,MeV, and
$L(\rho_{\mathrm{c}})=49.97$\,MeV. It should be mentioned that the
pressure of SNM is independent of the values of
$E_{\rm{sym}}(\rho_{\mathrm{c}})$ and $L(\rho_{\mathrm{c}})$. It is
seen from Fig.\,\ref{PressCorr} that the pressure of SNM $P(\rho)$
at $\rho=3\rho_0$ increases with $\rho_{0}$,
$M_{\mathrm{dirac}}^{\ast 0}$, $K_{0}$ and $J_{0}$ while decreases
with $E_{0}(\rho_{0})$. In particular, the pressure of SNM $P(\rho)$
at $\rho=3\rho_0$ displays a specially strong correlation with
$J_{0}$. We note that the pressure of SNM $P(\rho)$ at other
supra-saturation densities exhibits similar correlations with
$\rho_{0}$, $E_{0}(\rho_{0})$, $M_{\mathrm{dirac}}^{\ast 0}$,
$K_{0}$, and $J_{0}$. These features indicate that the pressure of
SNM at supra-saturation densities is sensitive to the $J_0$ value,
and thus the experimental constraints on the pressure of SNM at
supra-saturation densities may provide important information on the
$J_0$ value.

Since the pressure of SNM at supra-saturation densities is sensitive
to the $J_0$ value, the maximum mass $M_{\text{max}}$ of static
neutron stars is also expected to be sensitive to the $J_0$ value.
The mass and radius of static neutron stars can be obtained from
solving the Tolman-Oppenheimer-Volkoff (TOV) equations with a given
neutron star matter EOS. A neutron star generally contains core,
inner crust and outer crust from the center to surface. In this
work, for the core where the baryon density is larger than the
core-curst transition density $\rho_{\text{t}}$, we use the EOS of
$\beta$-stable and charge neutral npe$\mu$ matter obtained from the
nonlinear RMF model. In the inner crust with densities between
$\rho_{\text{out}}$ and $\rho _{\text{t}}$ where the nuclear pastas
may exist,  we construct its EOS (pressure $P$ as a function of
energy density $\mathcal{E}$) according to
$P=a+b\mathcal{E} ^{4/3}$ because of our poor knowledge about its
EOS from first principle\,(\cite{Hor03,XuJ09}).
The $\rho_{\text{out}}=2.46\times 10^{-4}$ fm$^{-3}$ is the density
separating the inner from the outer crust. The constants $a$ and $b$
are then easily determined by the pressure and energy density at
$\rho _{\text{t}}$ and $\rho _{\text{out}}$\,\cite{XuJ09}. In
this work, the $\rho _{\text{t}}$ is determined self-consistently
within the nonlinear RMF model using the thermodynamical method
(see, e.g., \cite{Cai12} for the details). In the outer crust with
$6.93\times 10^{-13}$ fm$^{-3}<\rho <\rho _{\text{out}}$, we use the
EOS of BPS\,(\cite{BPS71,Iida97}), and in the region of $4.73\times
10^{-15}$ fm$^{-3}<\rho <$$6.93\times 10^{-13}$ fm$^{-3}$ we use the
EOS of FMT\,(\cite{BPS71}).

Similarly as in Fig.\,\ref{PressCorr}, we plot in
Fig.\,\ref{MassCorr} the maximum mass $M_{\text{max}}$ of static
neutron stars from the nonlinear RMF model based on the FSUGold
interaction by varying individually $\rho _{0}$, $E_{0}(\rho_{0})$,
$M_{\mathrm{dirac}}^{\ast 0}$, $K_{0}$, $J_{0}$,
$E_{\rm{sym}}(\rho_{\mathrm{c}})$, and $L(\rho_{\mathrm{c}})$ within
their empirical uncertain ranges. Indeed, one can see that the
$M_{\text{max}}$ displays a very strong positive correlation with
the $J_0$ parameter. In addition, the $M_{\text{max}}$ exhibits weak
positive correlation with the $M_{\mathrm{dirac}}^{\ast 0}$ and
$K_{0}$, and weak negative correlation with the $\rho _{0}$ and
$E_{0}(\rho_{0})$. It is interesting to see that the
$M_{\text{max}}$ is essentially independent of the values of
$E_{\rm{sym}}(\rho_{\mathrm{c}})$ and $L(\rho_{\mathrm{c}})$,
implying that, in the nonlinear RMF model, the $M_{\text{max}}$ is
basically determined by the isoscalar part of the nuclear matter
EOS. Since the seven microscopic parameters change with
the macroscopic quantities, it is thus not surprising to see that the maximum mass
of a neutron star based on the FSUGold interaction by varying macroscopic quantities
may be totally different from the default one from FSUGold, which is about
$1.74 M_{\odot}$. These features indicate that the observed largest mass of
neutron stars may put important constraint on the $J_0$ value.
We would like to point out that
the interaction FSUGold is only used in Figs.~\ref{PressCorr}
and ~\ref{MassCorr} as a reference for the correlation analyses
and using other RMF interactions will not change our conclusion.

\begin{figure}
\begin{center}
\includegraphics[width=0.48\textwidth]{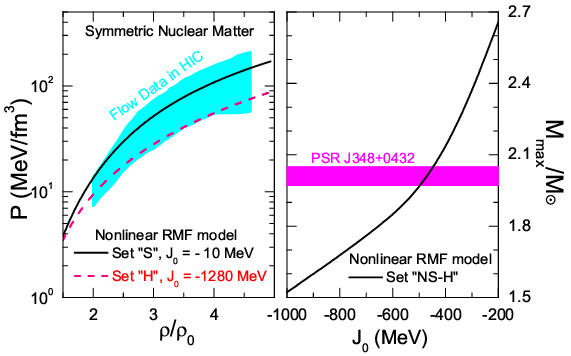}
\caption{\label{FlowNStar}  (Color online) Left window: Pressure of
SNM as a function of baryon density. The solid (dashed) line is the
prediction from the nonlinear RMF model with $J_0=-10$ ($-1280$) MeV
and the set ``S (H)" for $\rho_{0}$, $E_{0}(\rho_{0})$,
$M_{\mathrm{dirac}}^{\ast 0}$, and $K_{0}$. The band represents the
constraints from flow data in HIC\,\citep{Dan02}. Right window: The
maximum mass of static neutron stars as a function of $J_0$ in the
nonlinear RMF model with the set ``NS-H" for $\rho_{0}$,
$E_{0}(\rho_{0})$, $M_{\mathrm{dirac}}^{\ast 0}$, $K_{0}$,
$E_{\rm{sym}}(\rho_{\mathrm{c}})$ and $L(\rho_{\mathrm{c}})$. The
band represents mass $2.01\pm 0.04M_{\odot }$ for PSR
J0348+0432\,\citep{Ant13} .}
\end{center}
\end{figure}

Experimentally, the pressure of SNM at supra-saturation densities
(from $2\rho_0$ to about $5\rho_0$) has been constrained by
measurements of collective flows in HIC\,(\cite{Dan02}), which is
shown as a band in the left window of Fig.\,\ref{FlowNStar}. In the
nonlinear RMF model, if one only changes the $J_0$ value while the
other $6$ macroscopic quantities are kept at their values in the
FSUGold interaction, one can find that the $J_0$ value should be in
the range of $-985\,\mathrm{MeV}\leq J_0\leq -327\,\mathrm{MeV}$ to
be consistent with the flow data in HIC\,(\cite{Dan02}). However,
keeping the other $6$ macroscopic quantities at their values in the
FSUGold interaction is obviously a strong assumption because the
extraction of the $J_0$ value from the flow data in HIC will also
depends on the values of $\rho_{0}$, $E_{0}(\rho_{0})$,
$M_{\mathrm{dirac}}^{\ast 0}$, and $K_{0}$, which can be varied
within their empirical uncertain ranges. For the nonlinear RMF
model, we use in this work the following empirical uncertain ranges
for these macroscopic quantities, i.e.,
$\rho_0=0.153\pm0.008\,\mathrm{fm}^{-3}$,
$E_0(\rho_0)=-16.2\pm0.3\,\mathrm{MeV}$, $M_{\mathrm{dirac}}^{\ast
0}/M=0.61\pm0.04$, and $K_0=230\pm20\,\mathrm{MeV}$, which represent
the typical uncertain ranges known or predicted from different
interactions in the nonlinear RMF model\,(\cite{Che07}).

Based on the pressure of SNM constrained by flow data in
HIC\,(\cite{Dan02}), to extract the upper limit of the $J_0$ value,
one should use the values of $\rho_{0}$, $E_{0}(\rho_{0})$,
$M_{\mathrm{dirac}}^{\ast 0}$, and $K_{0}$ that make the resulting
pressure of SNM as small as possible when $J_0$ is fixed. This can
be obtained by using $\rho_0=0.145\,\mathrm{fm}^{-3}$,
$E_0(\rho_0)=-15.9\,\mathrm{MeV}$, ${M_{\rm{dirac}}^{\ast
0}}/{M}=0.57$, and $K_0=210\,\mathrm{MeV}$, denoted as set ``S",
since the pressure of SNM $P(\rho / \rho_0)$ at supra-saturation
densities increases with $\rho_0$, $M_{\mathrm{dirac}}^{\ast 0}$ and
$K_{0}$ while decreases with $E_{0}(\rho_{0})$ as shown in
Fig.\,\ref{PressCorr}. With the set ``S" for $\rho_{0}$,
$E_{0}(\rho_{0})$, $M_{\mathrm{dirac}}^{\ast 0}$, and $K_{0}$, one
can find the upper limit of $J_0= -10$ MeV for the $J_0$ value,
which is indicated by solid line in the left window of
Fig.\,\ref{FlowNStar}. For $J_0 > -10$ MeV, the model would
over-predict the pressure of SNM constrained by flow data in
HIC\,(\cite{Dan02}). Similarly, one can obtain the lower limit of
the $J_0$ value by using the values of $\rho_{0}$,
$E_{0}(\rho_{0})$, $M_{\mathrm{dirac}}^{\ast 0}$, and $K_{0}$ that
make the resulting pressure of SNM as large as possible when $J_0$
is fixed, and this can be obtained with
$\rho_0=0.161\,\mathrm{fm}^{-3}$, $E_0(\rho_0)=-16.5\,\mathrm{MeV}$,
${M_{\rm{dirac}}^{\ast 0}}/{M}=0.65$, and $K_0=250\,\mathrm{MeV}$,
denoted as set ``H". Using the set ``H" for $\rho_{0}$,
$E_{0}(\rho_{0})$, $M_{\mathrm{dirac}}^{\ast 0}$, and $K_{0}$, one
can extract the lower limit of $J_0= -1280$ MeV, which is indicated
by dashed line in the left window of Fig.\,\ref{FlowNStar}. The
model would under-predict the pressure of SNM constrained by flow
data in HIC\,(\cite{Dan02}) if $J_0 < -1280$ MeV. Therefore, from
the pressure of SNM constrained by flow data in HIC\,(\cite{Dan02}),
one can extract the constraint of $-1280 \text{ MeV} \le J_0 \le -10
\text{ MeV}$.

Recently, a new neutron star PSR J0348+0432 with a mass of
$2.01\pm0.04M_{\odot}$ was discovered\,(\cite{Ant13}), and this
neutron star is only the second pulsar with a precisely determined
mass around $2M_{\odot}$ after PSR J1614-2230\,(\cite{Dem10}) and
sets a new record of the maximum mass of neutron stars. The lower
mass limit of $1.97M_{\odot}$ for PSR J0348+0432 thus may set a
lower limit of the $J_0$ value below which the model cannot predict
a neutron star with mass equal or above $1.97M_{\odot}$. To extract
the lower limit of the $J_0$ value from the observed heaviest
neutron star PSR J0348+0432, one can use the values of $\rho_{0}$,
$E_{0}(\rho_{0})$, $M_{\mathrm{dirac}}^{\ast 0}$, $K_{0}$,
$E_{\rm{sym}}(\rho_{\mathrm{c}})$ and $L(\rho_{\mathrm{c}})$ that
make the resulting maximum mass of neutron stars as large as
possible when $J_0$ is fixed, and from Fig.\,\ref{MassCorr} this can
be obtained with $\rho_0=0.145\,\mathrm{fm}^{-3}$,
$E_0(\rho_0)=-16.5\,\mathrm{MeV}$, $K_0=250\,\mathrm{MeV}$,
${M_{\rm{dirac}}^{\ast 0}}/{M}=0.65$,
$E_{\rm{sym}}(\rho_{\mathrm{c}})=26\,\mathrm{MeV}$ and
$L(\rho_{\mathrm{c}})=60\,\mathrm{MeV}$, denoted as set ``NS-H".
This leads to a lower limit of $J_0= -494$ MeV for the $J_0$ value
as shown in the right window of Fig.\,\ref{FlowNStar} where the
maximum mass of neutron stars is plotted as a function of $J_0 $
when the other $6$ macroscopic quantities are fixed at their values
as in set ``NS-H". For $J_0 < -494$ MeV, the maximum mass of static
neutron stars predicted in the nonlinear RMF model would be always
smaller than $1.97M_{\odot}$. It should be pointed out that here the
interior of neutron stars has been assumed to be npe$\mu$ matter.
New degrees of freedom such as hyperons or/and quark matter that
could be present in the interior of neutron stars usually soften the
EOS of neutron star matter and thus a larger $J_0$ value would be
necessary to obtain a neutron star with mass of $1.97M_{\odot}$.
Therefore, including the new degrees of freedom in neutron stars
will be consistent with the constraint of $J_0 \ge -494$ MeV.

Combining the constraint of $-1280 \text{ MeV} \le J_0 \le -10
\text{ MeV}$ from the pressure of SNM constrained by flow data in
HIC\,(\cite{Dan02}) which favors a smaller $J_0$ value and the
constraint of $J_0 \ge -494$ MeV from the recently discovered
heaviest neutron star PSR J0348+0432\,(\cite{Ant13}) which favors a
larger $J_0$ value, one can extract the following constraint for the
$J_0$ parameter
\begin{equation}\label{cons_J0}
-494\,\mathrm{MeV}\leq J_0\leq -10\,\mathrm{MeV}.
\end{equation}
It should be emphasized that
the constraint $-494\,\mathrm{MeV}\leq J_0\leq -10\,\mathrm{MeV}$
represents a conservative extraction based on flow
data in HIC\,(\cite{Dan02}) and the recently discovered heaviest
neutron star PSR J0348+0432\,(\cite{Ant13}) in the nonlinear RMF
model. This is because we have not considered the possible
correlations existed among the $\rho_{0}$, $E_{0}(\rho_{0})$,
$M_{\mathrm{dirac}}^{\ast 0}$, $K_{0}$, $E_{\rm{sym}}(\rho_{\mathrm{c}})$
and $L(\rho_{\mathrm{c}})$, and have simply set simultaneously their
values in the boundary of their empirical uncertain ranges.
Considering the
correlations possibly existed among the $\rho_{0}$, $E_{0}(\rho_{0})$,
$M_{\mathrm{dirac}}^{\ast 0}$, $K_{0}$, $E_{\rm{sym}}(\rho_{\mathrm{c}})$
and $L(\rho_{\mathrm{c}})$ should further narrow the
constraint $-494\,\mathrm{MeV}\leq J_0\leq -10\,\mathrm{MeV}$,
and it will be interesting to see the quantitative constraint on the
$J_0$ parameter based on the data of finite nuclei, neutron stars, and heavy-ion
collisions using the exhaustive statistical analysis method although
this is beyond the scope of the present work.
The conservative constraint
$-494\,\mathrm{MeV}\leq J_0\leq -10\,\mathrm{MeV}$ obtained in the
present work indicates that, if the $J_0$ value is out of the
region $-494\,\mathrm{MeV}\leq J_0\leq -10\,\mathrm{MeV}$, the
nonlinear RMF model either cannot predict the pressure of SNM
constrained by flow data in HIC\,(\cite{Dan02}) or cannot describe
the recently discovered heaviest neutron star PSR
J0348+0432\,(\cite{Ant13}).
It is worth mentioning that the constraint on the $J_0$ depends on our knowledge of
the other six quantities. Any improvement on these six macroscopic quantities will
make the range for the $J_0$ constraint narrower. In addition, the extracted constraint on the $J_0$
could depend on the form of the energy density functional, and it will be interesting to see how
the constraint changes if other energy density functionals are used.

\begin{figure}
\begin{center}
\includegraphics[width=0.38\textwidth]{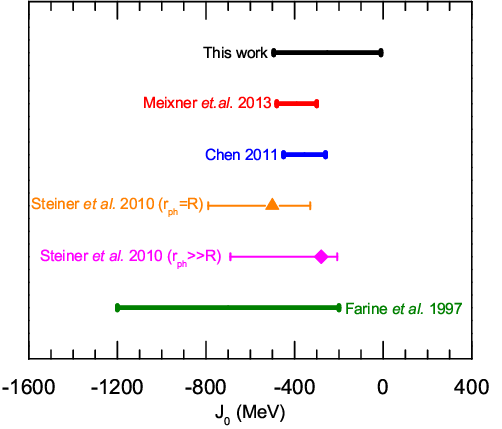}
\caption{\label{J0}  (Color online) Comparison between the
constraint of $J_0$ extracted in the present work and those from the
existing literature\,\citep{Far97,Ste10,Che11a,Mei13}.}
\end{center}
\end{figure}

In Fig.\,\ref{J0}, we show the comparison of the $J_0$ constraint
obtained in our analysis with those obtained with other analyses
and/or other methods\,(\cite{Far97,Ste10,Che11a,Mei13}), including
the constraint of $J_0 = -700 \pm 500$ MeV obtained by\,\cite{Far97}
from the analysis of nuclear GMR, the constraint of $J_0 =
-280^{+72}_{-410}$ ($-500^{+170}_{-290}$) MeV obtained
by\,\cite{Ste10} from analyzing a heterogeneous data set of six
neutron stars using a Markov chain Monte Carlo algorithm within a
Bayesian framework by assuming $r_{\text{ph}} \gg R$
($r_{\text{ph}}= R$) where $r_{\text{ph}}$ is the photospheric
radius at the time the flux is evaluated and $R$ is the stellar
radius, the constraint of $J_0 = -355 \pm 95$ MeV deduced
by\,\cite{Che11a} based on a correlation analysis method within SHF
energy density functional, and the constraint of $J_0 = -390 \pm 90$
MeV deduced by\,\cite{Mei13} who used the similar method
as\,\cite{Che11a}. It is seen that the constrained region of $J_0$
obtained in the present work has a remarkable overlap with those
existing in the literature. In particular, our present constraint
from the relativistic model is nicely consistent with the
constraints deduced from the non-relativistic SHF approach and they
all indicates that the $J_0$ parameter should be larger than about
$-500$ MeV.

\section{Summary}

Within the nonlinear relativistic mean field model, using macroscopic
nuclear matter characteristic parameters instead of the microscopic
coupling constants as direct input quantities, we have demonstrated that
the pressure of symmetric nuclear matter at supra-saturation densities
and the maximum mass of neutron stars provide useful probes for the
skewness coefficient $J_0$ of symmetric nuclear matter. In particular,
using the existing experimental constraints on the pressure of symmetric
nuclear matter at supra-saturation densities from flow data in heavy ion
collisions and the astrophysical observation of recently discovered
heaviest neutron star PSR J0348+0432, with the former requiring a smaller
$J_0$ while the latter a larger $J_0$, we have extracted a constraint of
$-494 \mathrm{MeV}\leq J_0\leq -10 \mathrm{MeV}$.

We have compared the
present constraint with the results obtained in other analyses, and found
they are nicely in agreement. In particular, our present constraint from
the relativistic model is nicely consistent with the constraints deduced
from the non-relativistic Skyrme-Hartree-Fock approach and they all
indicate that the $J_0$ parameter cannot be too small, namely, it should
be larger than about $-500$ MeV. The present constraint on the $J_0$
parameter provides important information on the high density behaviors of
the EOS of symmetric nuclear matter, and also may be potentially useful
for the determination of the high density behaviors of the EOS
of asymmetric nuclear matter, especially the high density symmetry energy.

\section*{Acknowledgments}
This work was supported in part by the Major State Basic Research
Development Program (973 Program) in China under Contract Nos.
2013CB834405 and 2015CB856904, the National Natural Science
Foundation of China under Grant Nos. 11625521, 11275125 and
11135011, the Program for Professor of Special Appointment (Eastern
Scholar) at Shanghai Institutions of Higher Learning, Key Laboratory
for Particle Physics, Astrophysics and Cosmology, Ministry of
Education, China, and the Science and Technology Commission of
Shanghai Municipality (11DZ2260700).


\begin{references}

\bibitem{Bla80} J.P. Blaizot, Phys. Rep. \textbf{64}, 171 (1980).

\bibitem{LiBA98} B.A. Li, C.M. Ko, and  W. Bauer, Int. J. Mod. Phys. E
\textbf{7}, 147 (1998).

\bibitem{Dan02} P. Danielewicz, R. Lacey, and W.G. Lynch, Science,
\textbf{298}, 1592 (2002).

\bibitem{Bar05} V. Baran, M. Colonna, V. Greco, and M. Di Toro, Phys. Rep. \textbf{410}, 335 (2005).

\bibitem{Ste05} A.W. Steiner, M. Prakash, J.M. Lattimer, and P.J.
Ellis, Phys. Rep. \textbf{411}, 325 (2005).

\bibitem{Che07a} L.W. Chen, C.M. Ko, B.A. Li, and G.C. Yong, Front.
Phys. China \textbf{2}, 327 (2007).

\bibitem{LCK08} B.A. Li, L.W. Chen, and C.M. Ko, Phys. Rep. \textbf{464}, 113 (2008).

\bibitem{Nat10} J. B. Natowitz et al., Phys. Rev. Lett. \textbf{104}, 202501 (2010).

\bibitem{Tsa12} B.M. Tsang \textit{et al.}, Phys. Rev. C \textbf{86}, 105803 (2012).

\bibitem{Tra12} W. Trautmann and H.H. Wolter, Int. J. Mod. Phys. E \textbf{21}, 1230003 (2012).

\bibitem{Che14} L.W. Chen, C.M. Ko, B.A. Li, C. Xu, and J. Xu, Eur. Phys. J. A \textbf{50}, 29 (2014).

\bibitem{Hor14} C.J. Horowitz, E.F. Brown, Y. Kim, W.G. Lynch, R. Michaels, A. Ono, J. Piekarewicz, M.B. Tsang, and H.H. Wolter, J. Phys. G \textbf{41}, 093001 (2014).
\bibitem{LiBA14} B.A. Li, A. Ramos, G. Verde, and I. Vidana, Eur. Phys. J. A \textbf{50}, 9 (2014).
\bibitem{Liu15} X. Q. Liu et al., Nucl. Sci. Tech. \textbf{26}, S20508 (2015); F. F. Duan et al., Nucl. Sci. Tech. \textbf{27}, 131 (2016)
\bibitem{Bal16} M. Baldo and G.F. Burgio, Prog. Part. Nucl. Phys. \textbf{91}, 203 (2016).
\bibitem{LiBA17} B.A. Li, Nucl. Phys. News, in press, (2017) [arXiv:1701.03564]

\bibitem{Gle00} N.K. Glendenning, Compact Stars, 2nd edition, Spinger-Verlag New York, Inc., 2000.

\bibitem{Lat04} J.M. Lattimer and M. Prakash, Science \textbf{304}, 536
(2004); Phys. Rep. \textbf{442}, 109 (2007).

\bibitem{Lat12} J.M. Lattimer, Annu. Rev. Nucl. Part. Sci. \textbf{62}, 485 (2012).

\bibitem{SNRev} K. Kotake, K. Sato, and K. Takahashi, Rep. Prog. Phys. \textbf{69},
971 (2006); H.-Th. Janka, K. Langanke, A. Marek, G. Pinedob, and B. M$\ddot{\mathrm{u}}$ller,
Phys. Rep. \textbf{442}, 38 (2007).

\bibitem{Hem12} M. Hempel, T. Fischer, J. Schaffner-Bielich, and M. Liebend$\ddot{\mathrm{o}}$rfer,
Astrophys. J. \textbf{748}, 70 (2012).

\bibitem{Mei13} M. Meixner, J.P. Olson, G. Mathews, N.Q. Lan, and H.E.
Dalhed, arXiv:1303.0064, (2013).

\bibitem{Oer17} M. Oertel, M. Hempel, T. Kl\"{a}hn, and S. Typel, Rev. Mod. Phys. \textbf{89}, 015007 (2017).

\bibitem{Che09} L.W. Chen, B.J. Cai, C.M. Ko, B.A. Li, C. Shen, and
J. Xu, Phys. Rev. C \textbf{80}, 014332 (2009).

\bibitem{Che11a} L.W. Chen, Sci. China: Phys. Mech. Astron. \textbf{54},
suppl. 1, s124 (2011) [arXiv:1101.2384].


\bibitem{You99} D.H. Youngblood, H.L. Clark, and Y.-W. Lui, Phys. Rev. Lett.
\textbf{82}, 691 (1999).

\bibitem{Shl06} S. Shlomo, V.M. Kolomietz, and G Col\`{o}, Eur. Phys. J. A
\textbf{30}, 23 (2006).

\bibitem{Col09} G. Col\`{o}, AIP Conf. Proc. \textbf{1128}, 59 (2009)
[arXiv:0902.3739].

\bibitem{Pie10} J. Piekarewicz, J. Phys. G \textbf{37}, 064038 (2010).

\bibitem{Che12} L.W. Chen and J.Z. Gu, J. Phys. G \textbf{39}, 035104 (2012).

\bibitem{Che12a} L.W. Chen, Nucl. Phys. Rev. 37, 273 (2014); arXiv:1212.0284.

\bibitem{Li12} B.A. Li, L.W. Chen, F.J. Fattoyev, W.G. Newton, and C. Xu,
J. Phys.: Conf. Ser. \textbf{413}, 012021 (2013) [arXiv:1212.1178].

\bibitem{Zha13} Z. Zhang and L.W. Chen, Phys. Lett. \textbf{B726}, 234 (2013).

\bibitem{Ser86} B.D. Serot and J.D. Walecka, Adv. Nucl. Phys. \textbf{16}, 1
(1986); Int. J. Mod. Phys. E \textbf{6}, 515 (1997).

\bibitem{Rei89} P.-G. Reinhard, Rep. Prog. Phys. \textbf{52}, 439 (1989).

\bibitem{Rin96} P. Ring, Prog. Part. Nucl. Phys. \textbf{37}, 193 (1996).

\bibitem{Men06} J. Meng, H. Toki, S.G. Zhou, S.Q. Zhang, W.H. Long, and L.S.
Geng, Prog. Part. Nucl. Phys. \textbf{57}, 470 (2006).

\bibitem{Sug94} Y. Sugahara and H. Toki, Nucl. Phys. \textbf{A579}, 557 (1994).

\bibitem{Ren96} Z.Z. Ren, Z.Y. Zhu, Y.H. Cai, and G.O. Xu, Phys. Lett. \textbf{B380}, 241 (1996).

\bibitem{Lal97} G.A. Lalazissis, J. K$\ddot{\rm{o}}$nig, and P. Ring, Phys. Rev. C \textbf{55}, 540 (1997).

\bibitem{Lon04} W.H. Long, J. Meng, N. Van Giai, and S.G. Zhou, Phys. Rev. C \textbf{69}, 034319 (2004).

\bibitem{Jia05} W.Z. Jiang, Z.Z. Ren, T.T. Wang, Y.L. Zhao, and Z.Y. Zhu, Eur. Phys. J. A 25, 29 (2005).

\bibitem{Jia10} W.Z. Jiang, Phys. Rev. C \textbf{81}, 044306 (2010).

\bibitem{Fat10} F.J. Fattoyev, C.J. Horowitz, J. Piekarewicz, and G. Shen, Phys. Rev. C \textbf{82},
055803 (2010).

\bibitem{Agr12} B.K. Agrawal, A. Sulaksono, and P.-G. Reinhard, Nucl. Phys. \textbf{A882}, 1 (2012).

\bibitem{Fat13} F.J. Fattoyev, J. Carvajal, W.G. Newton, and B.A. Li, Phys. Rev. C \textbf{87},
015806 (2013).

\bibitem{Mul96} H. M\"{u}ller and B. D. Serot, Nucl. Phys. \textbf{A606},
508 (1996).

\bibitem{Hor01} C.J. Horowitz, and J. Piekarewicz, Phys. Rev. Lett \textbf{%
86}, 5647 (2001); Phys. Rev. C \textbf{64}, 062802 (R) (2001); Phys. Rev. C
\textbf{66}, 055803 (2002).

\bibitem{Tod05} B.G. Todd-Rutel and J. Piekarewicz, Phys. Rev. Lett.
\textbf{95}, 122501 (2005).

\bibitem{Che07} L.W. Chen, C.M. Ko, and B.A. Li, Phys. Rev. C \textbf{76}, 054316
(2007).

\bibitem{Cai12} B.J. Cai and L.W. Chen, Phys. Rev. C \textbf{85},
024302 (2012).

\bibitem{Cai14} Equation~(\ref{J0_for}) was given in the first version of the present paper, i.e., arXiv:1402.4242v1 [nucl-th], in February, 2014.

\bibitem{Cai12a} B.J. Cai and L.W. Chen, Phys. Lett. \textbf{B711},
104 (2012).

\bibitem{Che10} L.W. Chen, C.M. Ko, B.A. Li, and J. Xu, Phys. Rev. C
\textbf{82}, 024321 (2010).

\bibitem{Zha14} Z. Zhang and L.W. Chen, Phys. Rev. C \textbf{90}, 064317 (2014).

\bibitem{Che11} L.W. Chen, Phys. Rev. C \textbf{83}, 044308 (2011).


\bibitem{Hor03} J. Carriere, C.J. Horowitz, and J. Piekarewicz, Astrophys.
J. \textbf{593}, 463 (2003).

\bibitem{XuJ09} J. Xu, L.W. Chen, B.A. Li, and H.R. Ma, Phys. Rev. C \textbf{%
79}, 035802 (2009); Astrophys. J. \textbf{697}, 1549 (2009).


\bibitem{BPS71} G. Baym, C. Pethick, and P. Sutherland, Astrophys. J. \textbf{%
170}, 299 (1971).

\bibitem{Iida97} K. Iida and K. Sato, Astrophys. J. \textbf{477}, 294 (1997).


\bibitem{Ant13} J. Antoniadis \textit{et al.}, Science \textbf{340}, 6131 (2013).

\bibitem{Dem10} P. Demorest, T. Pennucci, S. Ransom, M. Roberts, and J. Hessels, Nature \textbf{467}, 1081 (2010).



\bibitem{Far97} M. Farine, J.M. Pearson, and  F. Tondeur, Nucl. Phys. \textbf{A615}, 135 (1997).

\bibitem{Ste10} A.W. Steiner, J.M. Lattimer, and E.F. Brown, Astrophys. J. \textbf{722}, 33 (2010).

\end{references}
\end{document}